\documentclass[12pt]{article}
\usepackage{amsmath,amssymb}

\begin{document}

\begin{center}

\section*{Closed Vortex Filament in a Cylindrical Domain: Circulation Quantization}

{S.V. TALALOV}

{Department of Applied Mathematics, Togliatti State University, \\ 14 Belorusskaya str.,
 Tolyatti, Samara region, 445020 Russia.\\
svt\_19@mail.ru}

\end{center}

\begin{abstract}
 This article investigates quantum  oscillations of a vortex ring with zero thickness that evolves  in a cylindrical domain
$V = D \times [0,L]$.  The symbol $D$ denotes the planar domain which is bounded by  some closed connected  curve $S$.
The quantization scheme of this dynamical system is based on the approach proposed by the author earlier.
As result, we find the discrete  values $\Gamma_n$ for circulation $\Gamma$.
 In contrast to the traditional approach, where such quantities are usually postulated, the values $\Gamma_n$ are deduced  rigorously as the consequence of the
conventional scheme of quantum theory. The model  demonstrates the splitting of levels also.
In particular,  the levels correction values
  depend on the  domain $V$: both the cylinder height $L$ and the form of the curve  $S$ affect the final formula for the quantities  $\Gamma_n$.
Moreover, we prove that the basic  circulation levels demonstrate a "fine structure".  These anomalous terms, which are proportional to the value $\hbar^2$, are calculated in the article as well.
{  The conclusions are compared with some results of numerical simulations by other authors.}
\end{abstract}

{\bf keywords:}   vortex ring, constrained Hamiltonian systems,  circulation quantization

 {\bf PACS numbers:}   47.10.Df    47.32.C

\vspace{5mm}

   As it seems, the theory of quantum fluids cannot be considered complete.
Without attempting  to make a review of the literature on this topic, we will mention only the Landau two-fluids model (see, for example, \cite{Putt}) which marks the beginning of a whole new direction of research in this field.
Probably, the quantum description of vortex filaments arising in  fluids is 
 one of the most interesting areas for research. 
Thus,  a number of authors (see, for example, \cite{TsFuYu} ) assume that the topological defects  -- vortices --
are the key to  understanding    the turbulence nature in  quantum fluids.
{   In the recent  paper \cite{MuPoKr} it was emphasized  that the investigation of individual quantum vortices at certain scales is very important in this regard.
 In the work\cite{MuPoKr}  the numerical investigations were performed within the framework of a generalized Gross-Pitaevskii model.
The determining role of circulation in the study of turbulence was also pointed out in article \cite{ISY}.}

The problem of circulation quantization is inextricably linked with these studies.
So far,  the quantization  was postulated \cite{Putt}:

\begin{equation}
        \label{Circ_post}
				\Gamma_n \equiv  \oint_\gamma {\bf u}(\ell)d{\boldsymbol\ell}   =\frac{ n \hbar}{\mu}\,\qquad  n = 0,1,2,\dots,		
\end{equation}				
	where  $\gamma$ is  some closed curve in space $E_3$  and $	{\bf u}(\ell) \equiv {\bf u}( {\bf r}(\ell))$ -- fluid velocity  in the point 	${\bf r}(\ell)\in \gamma$.
	Parameter $\mu$ is some parameter with a dimension of  mass. 		
As  it has been repeatedly stated in the literature, such quantization rules are similar to the quantization rules in the old Bohr quantum theory.
In contradiction with the contemporary quantum mechanics,  quantum levels in the old theory  do not depend on boundary conditions.
Moreover, the quantization of each observable should be deduced from the general principles of quantum theory, and not be postulated separately.

In this article, we suggest the scheme of the quantum description of the vortex ring moving in the cylindrical domain.  
The approach under consideration is a further development of the ideas stated in the work \cite{Tal}.
The suggested method  leads to circulation quantization naturally.

   We consider the following dynamical system. A closed vortex filament  with zero thickness 
	described by the curve ${\boldsymbol r} = {\boldsymbol r}(t,\ell)$, 
 moves parallel to the axis  of the cylindrical fluid-filled  domain $V = D \times [0,L]$. 
{  Note that the quantum vortex structures  in a similar domain were investigated in the article \cite{InNaTs} recently.} 

 The dynamics of our object  is described by the Local Induction Equation.  
 Naturally, we  assume that the Z-axis coincides with the axis of the considered cylinder. 
The theory has three   dimensional constants: the fluid's density $\varrho_0$, the speed of sound in this fluid $v_0$ and the natural scale length 
$$R_0 \simeq \max\|{\boldsymbol r}_1 - {\boldsymbol r}_2\|\,, \qquad {\boldsymbol r}_1, {\boldsymbol r}_2 \in S\,.$$
For convenience, we will use the auxiliary constants  $t_0 = R_0/v_0$, $\mu_0 $ and   ${\cal E}_0 = \mu_0 v_0^2/2$. These constants  define the scales of time, masses and energy correspondingly. 
In the work \cite{Tal}, parameter $\mu_0 $ appeared as the central charge for the centrally extended Galilei group.
Despite the fact that there is a natural mass parameter $\tilde\mu_0 =  \varrho_0 R_0^2 L$  in this work,
we will use the mass $\mu_0 $ too. Therefore, we have an additional dimensionless parameter here:
$\alpha = \mu_0/\tilde\mu_0$. We will discuss its significance in theory later.

For  convenience, we  introduce 
 the  dimensionless parameter $\xi = \ell/R$  and   the  ''dimensionless time'' $\tau = t/t_0$ for all 
evolving  closed curves  ${\boldsymbol{r}}(t,\ell)$  with length  $2\pi R$. 
 With these denotations,     
the vortex filament we are considering is given by the formula
\begin{equation}
        \label{involve}
                                   {\boldsymbol{r}}(\tau,\xi) =  \boldsymbol{q} + 
   {R}\, \int\limits_{0}^{2\pi}  \left[\, {\xi - \eta}\,\right] {\boldsymbol j}(\tau,\eta) d\eta\,,  \qquad R << R_0\,,
                  \end{equation}

where the notation $[\,x\,]$ means the integer part of the number $x/{2\pi}$ and the  variables  $\boldsymbol{q} = \boldsymbol{q}(\tau)$  may be $\tau$-dependent.
$2\pi$-periodical   vector function ${\boldsymbol j}(\xi) \in E_3$ defines the unit tangent vector. 
 {  Because our  vortex filament is closed,  }    the following   equalities are fulfilled too:
            \begin{equation}
  \label{constr_j_0}
    \int\limits_{0}^{2\pi}{j}_k(\xi)\,d\xi  = 0\, \qquad  (k=x,y,z)\,,
   \end{equation}

 The 	function  ${\boldsymbol{r}}(\tau,\xi)$ satisfies the LIE equation
								\begin{equation}
        \label{LIE_eq}
        \partial_\tau {\boldsymbol{r}}(\tau ,\xi) = \frac{1}{R_0}\,
        \partial_\xi{\boldsymbol{r}}(\tau ,\xi)\times\partial_\xi^{\,2}{\boldsymbol{r}}(\tau ,\xi)\,. 
        \end{equation}

In addition to the equation (\ref{LIE_eq}) that describes the evolution of the  
 curve ${\boldsymbol{r}}(\cdot,\xi)$, we postulate the standard hydrodynamic formula
for the momentum \cite{Batche}:
	  
   \begin{equation}
        \label{p_and_m_st}
        \tilde{\boldsymbol{p}} = \frac{\varrho_0}{2 }\,\int\,\boldsymbol{r}\times\boldsymbol{\omega}(\boldsymbol{r})dV\,.
                 \end{equation}

The vector  $\boldsymbol{\omega}(\boldsymbol{r})$  means   vorticity.   As well-known { (see, for example \cite{AlKuOk})}, 
  the vorticity of the  closed vortex filament  is calculated by means of the formula 
   \begin{equation}
        \label{vort_w}
     \boldsymbol{\omega}(\boldsymbol{r}) =  \Gamma
                  \int\limits_{0}^{2\pi}\,\delta(\boldsymbol{r} - \boldsymbol{r}(\xi))\partial_\xi{\boldsymbol{r}}(\xi)d\xi\,,
       \end{equation}  
where  symbol 	$\Gamma$ means circulation. 
Taking into account the formulae  (\ref{involve}),   (\ref{constr_j_0}) and   (\ref{vort_w}),  
 the following  expression for the canonical momentum is deduced {  by direct calculations}:
	\begin{eqnarray}
        \label{impuls_def}
                 \tilde{\boldsymbol{p}} &  = &   \varrho_0 {R}^2 \Gamma                  {\boldsymbol f} \,, \\
   {\boldsymbol f}  &  = &   \frac{1}{2}\iint\limits_{0}^{2\pi}  \left[\, {\xi - \eta}\,\right]\,{\boldsymbol j}(\eta)\times{\boldsymbol j}(\xi)d\xi  d\eta\,.\nonumber
      \end{eqnarray}

    In the work \cite{Tal} the author investigated  the small  perturbations   
		of the  vortex ring  described by the fotmula (\ref{involve}) with tangent vector
\begin{equation}
        \label{tang_v}
                                   {\boldsymbol{j}_0}(\xi) = \{ - \sin\xi\,, \quad \cos\xi\,, \quad 0\}\,
\end{equation}
  and coordinates 
	$  \boldsymbol{q}_0 = q_x^0 \boldsymbol{e}_x + q_y^0 \boldsymbol{e}_y + (q_z^0 + c\tau) \boldsymbol{e}_z\,$,   where  $ q_{x,y,z}^0 = const\,.$
		This means that we consider the quantities 
		\begin{equation}
        \label{perturb}
		\boldsymbol{q}  =  \boldsymbol{q}_0 + \varepsilon\, \boldsymbol{q}_{pert}\,, \qquad \boldsymbol{j}(\tau,\xi)  = {\boldsymbol{j}_0}(\xi) + \varepsilon {\boldsymbol{j}_{pert}}(\tau,\xi)\,,
		\end{equation} 
		where small parameter $\varepsilon < 1$ and the values  $|\boldsymbol{q}_{pert}|$  and $|{\boldsymbol{j}_{pert}}|$ are restricted.
		{  Let us note here the work \cite{Majda_Bertozzi}, where, in particular, the system of  small-perturbed  straightforward vortex filaments with  certain interactions  were investigated.} 
		
	Next, we will not write the ${''pert''}$ index explicitly, hoping that this will not lead to misunderstandings. 
	The symmetry of the original object allows the introduction of cylindrical coordinates
	$(\rho,\phi,z)$.  Thus,  the function  ${\boldsymbol j}(\tau ,\xi)$ can be written in form
	${\boldsymbol j}(\tau ,\xi)   
  =   j_\rho(\tau,\xi){\boldsymbol{e}}_\rho  +    j_\phi^{\,0}( { \tau}, \xi) \boldsymbol{e}_\phi +    j_z(\tau ,\xi)  {\boldsymbol{e}_z} $
		{ 
	and the linearized equation (\ref{LIE_eq}) takes the following form in the cylindrical basis:
\begin{eqnarray}
            \partial_\tau {\boldsymbol{j}}(\tau ,\xi)  & = & \Bigl({{j}_z}(\tau ,\xi)  +   \partial_\xi^{\,2}{{j}_z}(\tau,\xi)\Bigr) {\boldsymbol{e}_\rho} - \nonumber\\ 
~ & - &  \Bigl(   \partial_\xi^{\,2}{{j}_\rho}(\tau,\xi)  -2\, \partial_\xi {j}_\phi (\tau,\xi)  \Bigr) \boldsymbol{e}_z  \,.\nonumber
\end{eqnarray}
This equation demonstrates  that $ \partial_\tau {j^0_\phi}(\tau ,\xi) \equiv 0\,.$ }

	Further  
	we will consider the case  $j_\phi^{\,0}({ \tau},\xi) \equiv 0 $ only and the complex-valued function
	$  {\sf j}  =  j_\rho  + i  j_z\,. $
	Here it is appropriate  to  introduce   the separate  notation ${\sf j}$  for the complex perturbation amplitude to avoid any ambiguity.  
	Linearized equation for the amplitude ${\sf j}(\tau,\xi)$ was deduced in the article \cite{Tal}:
	{ 
	\begin{equation}
        \partial_\tau {\sf j} =  - i  \partial_\xi^{\,2} {\sf j}    - \frac{i}{2}\Bigl({\sf j}   - \overline{\,\sf j\,}\, \Bigr)   \,.\nonumber
\end{equation}}
	
	After solving this equation, we have the following representation for the function ${\sf j}(\tau,\xi)$:

\begin{equation}
        \label{sol_1}
				{\sf j}(\tau ,\xi) = \sum_{n} {\sf j}_{\,n}\, e^{\,i\,[n\xi  +  n \sqrt{n^2 - 1}\,\tau\,]} \,,   
\end{equation}			

where ${\sf j}_{\,n} \equiv const$ and the coefficients $ \overline{\,	{\sf j}\,}_{\,-n}$  and  ${\sf j}_{\,n}$ are connected to each other as follows:

\begin{equation}
        \label{jn_jn-conj}
			\overline{\,	{\sf j}\,}_{\,-n}     =    2 \left[n\sqrt{n^2 -1} - n^2 + \frac{1}{2} \right]  {\sf j}_{\,n} \,.
	\end{equation}			

{  Restrictions  (\ref{constr_j_0})   lead to the restrictions for the coefficients ${\sf j}_{\,0}$ and  ${\sf j}_{\pm 1}$:
			\[ {\sf j}_{\,0}  =  \,\overline{\,{\sf j}\,}_{\,0}\,, \qquad {\sf j}_{\, 1}  =  - \overline{\,{\sf j}\,}_{\,-1}\,.\]
						Although these restrictions are deduced from  the constraint (\ref{constr_j_0}), they are consistent with the formula (\ref{jn_jn-conj}). }

The formulas (\ref{impuls_def}),    (\ref{sol_1}) and (\ref{constr_j_0}) reveal  the following connection between the perturbation  ${\sf p} = p_x + i p_y$  of the momentum  (\ref{p_and_m_st})
and the amplitude ${\sf j}$:
\begin{equation}
        \label{p_pert_classic}
								\tilde{\sf p} = 2\pi\varrho_0 R^2\Gamma\, {{\sf j}\,}_{\,-1}\,. 
				\end{equation}	
				
				The projection  $p_z$ of the momentum  on the Z axis turns out to be unperturbed and is written as follows:										
				\begin{equation}
        \label{pz_classic}
								{{\tilde p}_z} = \pi\varrho_0 R^2\Gamma\, . 
				\end{equation}	
				
				Proofs and details can be found in the work		\cite{Tal}.

It is also necessary to pay attention to the following important point.
The local variable ''fluid velocity'' is absent in our theory.	
However, we intend  take  the movement of the fluid that surrounds the vortex filament into account,  in a certain way.
To do this, we  declare the value $\Gamma$ as a dynamic variable here,
 in addition to  the variables $\boldsymbol{j}(\xi)$ and  $\boldsymbol{q}$.
We believe that this assumption allows us  to take into account the dynamics of the surrounding fluid in a minimal way.
								We  denote as ${\mathcal A}$ the  set of the dynamical variables  
 $\{\,\boldsymbol{q}\,,{\boldsymbol j}(\xi)\,,\Gamma\,\}$.  
{   Despite the fact that the variable $\Gamma$ is a conserved quantity, such an extension of the original dynamical system (\ref{LIE_eq}) turns out to be nontrivial.
In order to make sure of this, }
 we now extend the set ${\mathcal A}$. 
 Let us denote as  ${\mathcal A}^{\,\prime}$ the set of the independent variables
	$({\boldsymbol q}\,;\, {\boldsymbol p}\,; {\boldsymbol j}(\xi)\,)$. 
	Formula   ${\boldsymbol p}  = \alpha {\tilde{\boldsymbol p}}$ and formula	
	(\ref{impuls_def}) perform the injection ${\cal F}$
		\[{\cal F}:\quad {\mathcal A} \quad \mathrel{\mathop{\longrightarrow}^{{\cal F}} } \quad {\mathcal A}^{\,\prime}\,, 
	\qquad  {\rm Ran}\, {\cal F} \equiv \Omega \subset {\mathcal A}^{\,\prime}\,  \]
	for every constant $\alpha$.  Next, we will demonstrate that  the set ${\mathcal A}^{\,\prime}$ is more appropriate 
		for subsequent quantization of our dynamical system.

It is clear that the formula (\ref{impuls_def}), which defines the momentum  through the variables of the set ${\mathcal A}$, 
is not  executed on the set ${\mathcal A}^{\,\prime}$ in general.  
Indeed, the vectors ${\boldsymbol p}$  and  ${\boldsymbol f}$ are independent vectors here.
But we can state that $\forall\,{\boldsymbol p}\in{\mathcal A}^{\,\prime}$,  $\forall\,{\boldsymbol j}\in{\mathcal A}^{\,\prime}$ the relation
\begin{equation}
        \label{impuls_A_prime}
              {\boldsymbol{p}}  =    \alpha\varrho_0 {R}^2 \Gamma \,  {\sf B}                 {\boldsymbol f} \,,\qquad  {\boldsymbol f} = {\boldsymbol f}[{\boldsymbol j}] 
   \end{equation} 
	 will be fulfilled 
	for some real number $\Gamma$ and some matrix ${\sf B} = {\sf B}(\theta) \in SO(3)$. 
	The angle $\theta$ is some angle  of rotation around the vector ${\boldsymbol n} = {\boldsymbol p} \times {\boldsymbol f} $.
	We can introduce certain  constraints which define subset $\Omega  \subset {\mathcal A}^{\,\prime} $ so that one-to-one correspondence 
 $ {\mathcal A} \longleftrightarrow  \Omega\,$  is established. 
For example, we can postulate the equality
\begin{equation}
        \label{con_mph}
 ( {\boldsymbol{p}} {\boldsymbol f})^2   =  {\boldsymbol{p}}^2 {\boldsymbol f}^2 \,, \qquad  {\boldsymbol f} = {\boldsymbol f}[{\boldsymbol j}]   \,, 
\qquad \,{\boldsymbol{p}}, {\boldsymbol j} \in {\mathcal A}^{\,\prime}\,.
\end{equation} 
{  Indeed, constraint (\ref{con_mph}) is fulfilled identical if and only if when the
 proportionality 
$ {\boldsymbol{p}} \propto {\boldsymbol f}$  takes place.  } 
 Let us note that in this paper we are investigating the circulation only. 
Despite the fact that the value $\Gamma$ is defined initially only on the set $ {\mathcal A} \approx  \Omega\,$, 
we can assume that it is also defined everywhere on the set ${\mathcal A}^{\,\prime} $.
Formula (\ref{impuls_A_prime}) defines the appropriate expansion  $\Omega \to {\mathcal A}^{\,\prime}$.
Moreover, the value $\Gamma$ is same both on the set $\Omega$ and the set  $ {\mathcal A}^{\,\prime}/ \Omega$ because the
identity ${\sf B}^{\sf T} {\sf B} = I$, $\forall\,{\sf B} \in SO(3)$. Indeed,
the following equality is fulfilled on the set  $ {\mathcal A}^{\,\prime}$: 
\begin{equation}
              \exists\, \Gamma\in {\sf R}\,: \qquad          |{\boldsymbol{p}}|^2  =   \alpha^2 \varrho_0^2 {R}^4 \Gamma^2 \,|{\boldsymbol f}|^2 \,.\nonumber
   \end{equation} 
Taking into account the equalities (\ref{p_pert_classic}) and         (\ref{pz_classic}),  let us write  this  formula in the following form:
				\begin{equation}
        \label{ness_cond}
				\exists\, \Gamma\in {\sf R}\,: \qquad {\Phi}_{\sf\Gamma} \equiv 
				|\,{\boldsymbol{p}}|^2 - \alpha^2\pi^2 \varrho_0^2 \,\Gamma^2 R^4\Bigl(1 + 4 \varepsilon^2 |\,{{\sf j}\,}_{\,-1}|^2   \Bigr) = 0\,. 
				\end{equation}
								The value $|\,{\boldsymbol{p}}|^2 = p_z^2 +  \varepsilon^2 |\,{\sf p}|^2$  here.  
								In fact, equality  ${\Phi}_{\sf\Gamma} = 0$ defines  the function  $ \Gamma = \Gamma ({\boldsymbol{p}},{\boldsymbol{j}})$ implicitly. 														
	This form  will be useful 				for our subsequent studies.

In order to obtain possible discrete values of circulation $\Gamma_n$, we need a quantum version of the model under consideration.
Both the hamiltonian structure of  the considered dynamical system  and subsequent  quantization scheme 
 have been considered in the work \cite{Tal} in detail\footnote{ The case  $V = {\sf R}_3$  was  considered  in the work \cite{Tal}   only. As a consequence,
both energy of the system and circulation take continuous values for this case. }.
 {  Without going into details in this paper , we  define   the fundamental hamiltonian variables here  as 
$ p_i\,, q_j\,, {\sf j}_{\,m}, \overline{\,\sf j\,}_{\,n} $  and their Poisson brackets:
\begin{eqnarray}
  \{p_i\,,q_j\} & = & \delta_{ij}\,,\qquad i,j = x,y,z\,, \nonumber \\
  \label{ja_jb}
  \{ {\sf j}_{\,m}, \overline{\,\sf j\,}_{\,n}\} & = & (i/{\mathcal E}_0 t_0)\, \delta_{mn}\,, \qquad m,n = -1,-2,\dots \nonumber
  \end{eqnarray}
		As a consequence,} 
 the structure of the Hilbert space $\boldsymbol{H}$   of  quantum states of our dynamical system can be defined as follows:
					\begin{equation}
	\boldsymbol{H}  =  \boldsymbol{H}_3 \otimes   \boldsymbol{H}_F\,,\nonumber
	\end{equation}
			where the symbol   $\boldsymbol{H}_3$  denotes the Hilbert space  of a free structureless particle in the domain $V$
			 (the space 	${\sf L}^2({ V}) = {\sf L}^2({ [0,L]}) \otimes {\sf L}^2({ D})$ in our case) 
			and  the symbol $\boldsymbol{H}_F $ denotes the Fock space for the infinite number of the harmonic oscillators.		 
The creation and annihilation operators which are defined in the space 	$\boldsymbol{H}_F $, have standard commutation relations
				\[ [\,\hat{a}_m, \hat{a}_n^+] = \hat{I}_F\,, \qquad \hat{a}_m|\,0\rangle = 0  \,, \qquad  m,n = 1,2,\dots \,, \qquad    |\,0\rangle \in    \boldsymbol{H}_F   \,,\]  

The variables ${\boldsymbol q}$,  ${\boldsymbol p}$ and ${\sf j}(\xi)$ are quantized in accordance with  the rule
\[ q_{x,y,z} \to  q_{x,y,z}\otimes \,\hat{I}_F  \,,\qquad  p_{x,y,z} \to - i\hbar\frac{\partial}{\partial q_{x,y,z}} \otimes \,\hat{I}_F \,,\]
				\[{\sf j}_{\,-n} \to \sqrt{\frac{\hbar}{t_0{\mathcal E}_0}}\, (\hat{I}_3 \otimes\,\hat{a}_n)\,, \]
		
					where  $ n = 1,2,\dots\, $.  The symbols   $\hat{I}_3$   and  $\hat{I}_F$  denote the identical operators in  spaces  $\boldsymbol{H}_3$ and $\boldsymbol{H}_F$ correspondingly.
Next, we will not write both the index $n = 1$ and the constructions $(\dots  \otimes \,\hat{I}_F)$,   $ (\hat{I}_3 \otimes\,\dots)$
 explicitly: $\hat{a}_{1}   =  \hat{a}$,   $(\hat{I}_3 \otimes\,\hat{a}_n) = \hat{a}_n$     and so on.

As  discussed above, in classical theory there should be constraints on the set 
${\mathcal A}^{\,\prime} $,
  to ensure the parallelism of vectors  $\boldsymbol{p}$ and $\boldsymbol{f}$.
Does this requirement make sense for the quantum version of the system in question?
First, there are no quantum states in the ''box'' $V$ which correspond to the definite momentum $\boldsymbol{p}$.
Moreover, the average square distance $\delta p^{\,2}$ of any component $p_{x,y,z}$ is limited from below.
Indeed,  the following inequalities take place for obvious geometrical reasons:
\[ \delta q_{x,y} \le R_0\,,\qquad   \delta q_{z} \le L\,,  \]
where  $\delta q = \sqrt{(q\, - \langle q\rangle)^2} $.
Taking into account Heisenberg's uncertainty principle, we can conclude that the  following inequalities take place for our quantum system:
\[ \delta p_{x,y} \ge \frac{\hbar}{2 R_0}\,,\qquad   \delta p_{z} \ge\frac{\hbar}{2 L}\,.  \]
On the contrary,  operator $\hat{\boldsymbol{f}}$ has states with  certain eigenvalues $\boldsymbol{f}$.
These conclusions make possible to consider space $\boldsymbol{H}$  completely, not restricted  to any subspace which is the consequence of classical constraints.
In fact, we quantize the extended classical system ${\mathcal A}^{\,\prime}$ instead of the initial system ${\mathcal A}$. 
We have demonstrated above that the value $\Gamma$ we are interested in is the same for the set ${\mathcal A}^{\,\prime}$ and the constraint surface 
$\Omega \subset {\mathcal A}^{\,\prime}$.

 To find permissible circulation values, we consider the equation
\[  \widehat{\Phi}_{\sf\Gamma}|\psi_{\Gamma}\rangle = 0\,,\]
where the operator $\widehat{\Phi}_{\sf\Gamma}$ is the quantized function ${\Phi}_{\sf\Gamma}({\boldsymbol p}, {\sf j})$ (see (\ref{ness_cond})).
Explicitly, this equation is written as:
\begin{equation}
	\label{main_eq}
\left[-\hbar^2\biggl(\frac{\partial^2}{\partial z^2} + \varepsilon^2 \Delta_2 \biggr) -
 \alpha^2 \pi^2 \varrho_0^2 \,\Gamma^2 R^4\biggl(1 +   \varepsilon^2 \frac{8\,\hbar}{\mu_0  v_0 R_0}\, {\hat a}^+ {\hat a}  \biggr)\right]  |\psi_{\Gamma}\rangle = 0\,.  
\end{equation}
We will look for solutions to this equation in the form 
\[  |\psi_{\Gamma}\rangle =    |\Psi_1\rangle |\Psi_D\rangle|j\rangle\,,\] 
where
\[ |\Psi_1\rangle \in {\sf L}^2([0,L])\,, \quad 
|\Psi_D\rangle \in {\sf L}^2(D)\,, \quad |j\rangle \in \boldsymbol{H}_F\,.\]
Let us consider the vectors  $|\Psi_1\rangle = |\Psi^n_1\rangle $ and      $|\Psi_D\rangle = |\Psi^m_D\rangle $ such that 
\[\langle q_z |\Psi^n_1\rangle \equiv \Psi_n(q_z) = \sin\Bigl(\frac{\pi n}{L} \Bigr)\,, \qquad \langle q_x, q_y |\Psi^m_D\rangle \equiv \Psi_m(q_x,q_y)\,,\]
in the coordinate representation.  Functions $\Psi_m(q_x,q_y)$ are eigenfunctions for the Laplace operator $\Delta_2$ in the planar domain $D$. These functions  satisfy the 
boundary conditions $\Psi_m(q_x,q_y)  =0 $ on the curve $S = \partial D$.
Next, we consider the vectors
\[ |j\rangle = |j_k\rangle \equiv  \frac{1}{\sqrt{k}}\,({\hat a}^+)^k|j_0\rangle\,,\qquad  |j_0\rangle = 
\hat{a}^+_{n_1}\hat{a}^+_{n_2}\dots\hat{a}^+_{n_{\ell}}| 0  \rangle \,, \quad  \qquad  n_j > 1 \,.  \] 

Choosing the solution of the equation  (\ref{main_eq})  in the form
\[ |\psi_{\Gamma}\rangle =  |n,m,k\rangle \equiv  |\Psi^n_1\rangle|\Psi^m_D\rangle |j_k\rangle\,,   \] 
we come to equality
\begin{equation}
	\label{equal_1}
\hbar^2\left[ \Bigl(\frac{\pi n}{L} \Bigr)^2 + \varepsilon^2 \lambda_m^2  \right]  -
 \pi^2\alpha^2 \varrho_0^2 \,\Gamma^2 R^4 \biggl(1+  \varepsilon^2 \frac{8\,\hbar\, k}{\mu_0  v_0 R_0}\, \biggr) = 0\,.
\end{equation}
Numbers $n,m,k$ are natural numbers here and  symbols $\lambda_m $ denote the eigenvalues of the Laplace operator $\Delta_2$ in the  domain $D$.
The equality (\ref{equal_1})  allows us to find the circulation  $\Gamma = \Gamma_{n,m,k} $ explicitly.
In accordance with our initial supposition (\ref{perturb}),
we can consider only such quantum numbers $n$, $m$, $k$ that ensure the fulfillment of the relations
\[ \lambda_m \le ({\pi n}/{L})\,, \qquad {8\hbar\, k} \le {\mu_0  v_0 R_0}\,. \] 
As a consequence, we can write an expression for circulation $\Gamma_{n,m,k} $
 as a series by constants $\varepsilon$ and $\hbar$: 
\begin{equation}
	\label{G_final}
 \Gamma_{n,m,k} = \frac{\hbar n}{\mu_v}\left[1 + \varepsilon^2 \biggl( \frac{1}{2}\Bigl(\frac{L \lambda_m}{\pi n}\Bigr)^2 - 
\frac{4 \hbar\, k}{\mu_0 v_0 R_0}     \biggr)  \right]  +  {\cal O}(\varepsilon^4, \hbar^3)\,, 
\end{equation}
where  the notation $\mu_v = \mu_0 (R/ R_0)^2$ was introduced.
The vectors $|n,m,k\rangle  $ form  the spaces $\boldsymbol{H_{n,m,k}} \in \boldsymbol{H}$ in a way that the following decomposition is fulfilled: 
\[  \boldsymbol{H}   =   \bigoplus_{m,n,k}  \boldsymbol{H_{n,m,k}}\,.       \]
The theory under consideration should be enhanced by the    ''superselection rules'': any pure state of the system  that is a linear combination of vectors from different spaces
$\boldsymbol{H_{n,m,k}}$ must be forbidden.

In the limit $\varepsilon \to 0$, 
 provided the quantum numbers  $m$ and $k$ are limited, we have the expression 
\[ \Gamma_{n,m,k}  \longrightarrow  \Gamma_{n} =  \frac{\hbar\, n}{\mu_v}\,,    \qquad n =0,1,2,\dots \]
 which coincides with the standard formula (\ref{Circ_post}).  We emphasize that this formula is  a consequence of the first principles of quantum theory here. 
 Apparently,  constant $\mu_0$ can be chosen based on agreement with a possible experiments.

Let us consider  terms of the formula (\ref{G_final}) in  square  brackets that are proportional to the value $\varepsilon^2$.  
The first term can be considered as  a certain ''form - factor'', because it is determined only by the form of the domain $V$.
The second term demonstrates the ''fine structure'' of the levels $\Gamma_{n,m,k}$  that was found. Indeed, it is proportional to the value $\hbar^2$.
Note also that the second term does not depend on the form  of the domain $V$ but depends on the excitations of the initial vortex ring.
{  What does formula (\ref{G_final}) predict for the value  $\mu_v\Gamma/\hbar$ in its experimental determination?   
It is clear that we have the ''large'' peaks in the integer values and certain ''small'' peaks at the non-integer values. 
 It will be interesting to note that in the work \cite{MuPoKr}  the numerical simulations for circulation 
confirms this distribution of the value of $\mu_v\Gamma/\hbar$.  
 Although this was explained by errors there, the model considered here gives a different explanation of this fact.}

\end{document}